\documentclass[12pt,preprint]{aastex}
\def\M21{\hbox{Mrk\,421} } 
 
\def\etal{et al.} 
 
\def\me{m_{\rm e}}

\def\ltsima{$\; \buildrel < \over \sim \;$} 
\def\simlt{\lower.5ex\hbox{\ltsima}} 
\def\gtsima{$\; \buildrel > \over \sim \;$}
\def\simgt{\lower.5ex\hbox{\gtsima}} 
\newcommand{\eqref}[1]{(\ref{#1})}

\shorttitle{Annomalous shear layers in relativistic jets}
\shortauthors{Aloy \& Mimica}

\begin{document}

\title{Observational Effects of Anomalous Boundary Layers in
  Relativistic Jets}

\author{M.A.\,Aloy\altaffilmark{1} 
  and P.\,Mimica\altaffilmark{1}}
%
\affil{Departamento de Astronom\'{\i}a y Astrof\'{\i}sica, 
           Universidad de Valencia, 46100 Burjassot, Spain}
\email{MAA, e-mail: Miguel.A.Aloy@uv.es}
\begin{abstract}
  Recent theoretical work has pointed out that the transition layer
  between a jet an the medium surrounding it may be more complex than
  previously thought. Under physically realizable conditions, the
  transverse profile of the Lorentz factor in the boundary layer can
  be non-monotonic, displaying the absolute maximum where the flow is
  faster than at the jet spine, followed by an steep fall
  off. Likewise, the rest-mass density, reaches an absolute minimum
  (coincident with the maximum in Lorentz factor) and then grows until
  it reaches the external medium value. Such a behaviour is in
  contrast to the standard monotonic decline of the Lorentz factor
  (from a maximum value at the jet central spine) and the
  corresponding increase of the rest-mass density (from the minimum
  reached at the jet core).  We study the emission properties of the
  aforementioned anomalous shear layer structures in kiloparsec-scale
  jets aiming to show observable differences with respect to
  conventional monotonic and smooth boundary layers.
\end{abstract}


\section{Introduction} 
  \label{sec:intro}

  The interaction of extragalactic jets with their environment leads,
  under rather general circumstances, to the stratification of the
  beam of the jet in the direction normal to its velocity. The
  morphology of FR I sources (e.g., M87, \citealp{owe89}; 3C 31,
  \citealp{lai96}; 3C 296, \citealp{hac97}) has been explained in
  terms of a jet whose dynamics is dominated by a boundary shear
  layer. In such layers the emissivity in radio and optical maps peaks
  and, in some cases (e.g., Mrk 501; \citealp{caw93}), the spectrum is
  rather flat, suggesting that the acceleration of non-thermal
  particles takes place right at the boundary region.  Further
  evidence in favor of such stratification is provided by radio
  polarization measurements, which indicate that, towards the jet
  boundary, the magnetic field is highly ordered and parallel to the
  jet axis \citep{per99}.

  A radial stratification of jets in FR~II radio galaxies has also
  been observed \citep{SBB98,caw93}.  The edge brightening found in
  3C353 has been interpreted by \cite{SBB98} as due to the Doppler
  hiding of the emission of the central spine of the jet,
  suggesting that most of the observed radiation comes from the jet
  boundary layer. On the other hand, the {\it rails of low
    polarization} found by \cite{SBB98} close the the jet boundary
  indicate that the magnetic field in the layer is either axial or
  toroidal but not radial. Even at parsec scale, the FR~II radio jet
  of 1055+018 exhibits a transverse structure consistent with a {\it
    spine - shear boundary layer} jet morphology \citep{ARW99}.

  Also sources which are in the borderline dividing FR~I and FR~II
  sources show evidences of radial jet stratification. A couple of
  examples of such sources are 3C~15 and Cen~A.  In the first case the
  jet is generally narrower in the optical than in the radio
  \citep{Dulwichetal07}, and simple spine-sheath model may account for
  the polarization angle differences seen in the optical and radio
  data. For Cen~A, the recent X-ray data \citep{Kataokaetal06} can
  also be properly explained by a stratified jet model with a radially
  decreasing velocity field.

  The aforementioned observational evidence along with some others
  (e.g., \citealp{chi00}), suggests that the velocity in some jet
  boundary layers is smaller (but still relativistic) than the
  velocity of the beam itself. On the other hand, laboratory
  experiments show that, almost unavoidably, turbulent interaction
  layers may develop as jets propagate into a viscous medium (e.g.,
  \citealt{BR74}). These laboratory shear layers display a radial
  velocity profile roughly monotonic in which the velocity of the jet
  core smoothly decreases until it vanishes at the external medium.

  From a theoretical point of view, the Reynolds number of the jet
  flow may reach values of $\sim 10^{23}$ and, hence, it seems
  unquestionable that such flows will quickly develop a turbulent
  boundary layer which spreads into the flow and leads to the
  entrainment and acceleration of ambient gas
  \citep{DeYoung93}. However, the nature and the amount of viscosity
  in relativistic jets is still largely unknown. Most probably, the
  effective viscosity in the lateral interaction between the jet an
  the external medium is a mixture of the turbulent eddy viscosity,
  magnetic viscosity (based on the finite size of the Larmor radius;
  see, e.g., \citealp{Baan80}) and {\it cosmic-ray} viscosity (i.e.,
  the viscosity originated due to the diffusion of the momentum
  carried by energetic particles; see, e.g., \citealp{EJM88}).
  Whichever the origin of the viscosity is, the boundary layer itself
  results from the nonlinear growth of Kelvin-Helmholtz (KH)
  instabilities on the jet surface. The fact that relativistic jets
  are prone to these instabilities was demonstrated theoretically in
  the pioneer work of \cite{TS76}. Later, many other papers have
  explored the stability of relativistic jets with respect to KH
  instabilities in the vortex-sheet approximation (e.g.,
  \citealt{FTZ78,Hardee79}). \cite{FTZ82} and \citet{bir91} presented
  the first attempts to study (in the linear regime) the influence of
  a finite thickness shear layer in classical and relativistic jets,
  respectively. Very recently, \cite{Hardee07} has also considered the
  stability properties (in the linear regime) of spine-sheath
  magnetized, relativistic jets in which the magnetic field is
  parallel to the flow velocity. The development of KH modes in
  sheared relativistic (slab) flows from the linear (analytic) regime
  to the non-linear regime by means of relativistic hydrodynamic
  simulations has been carried forward by
  \cite{Peruchoetal07}. Alternative approaches to the linear analysis
  of the stability of stratified relativistic jets and sheared
  relativistic jets have been performed by \cite{HS96}, \cite{Urpin02}
  \cite{alo02} and \cite{MK07}, respectively.

  Further support in favor of radial jet stratification as a
  consequence of the interaction between a jet and its environment is
  provided by three-dimensional hydrodynamic simulations of
  relativistic, large scale jets \citep{alo99a,alo99b,alo99c,alo00}
  and two-dimensional simulations of relativistic magnetized jets
  \citep{LA05}. These simulations show that radially stratified jets
  are developed out of initially uniform beams. These numerical works
  display also a rather smooth transition layer with decreasing
  transverse velocity towards the jet boundary and high specific
  internal energy. It must be noted that the viscosity in such
  simulations is of numerical origin, and mimics only qualitatively
  actual viscous flows.

  In addition to the existence of {\it environmental} reasons to
  produce a radial flow stratification in astrophysical jets, this
  effect can also be a natural consequence of the jet formation
  process. Jet launching from accretion disks may directly lead to a
  certain mass flux profile and/or magnetic flux profile within the
  jet. This possibility has been verified by means of axisymmetric MHD
  simulations both in the classical \citep{PCO06,Fendt06}, as well as
  in the general relativistic \citep{MN07} regime.

  \cite{AR06} show that, under physically plausible conditions (see
  Sect.~\ref{sec:RMHD_boost}), the lateral structure of the
  interaction layer of a relativistic flow can be richer than
  previously thought. This is also the case in relativistic magnetized
  jets \citep{Mizuetal08}. Due to a purely relativistic phenomenon
  (the conversion of specific enthalpy into bulk Lorentz factor across
  a flow discontinuity), the jet velocity may increase right at the
  contact discontinuity (CD) separating the jet and the external
  medium. Beyond the CD the velocity decreases steeply to zero in the
  radial direction. The growth of the velocity or, equivalently, of
  the Lorentz factor of the beam is associated to the development of a
  rarefaction wave that emerges form the CD where the density and the
  pressure decrease with respect to their corresponding values at the
  jet spine. The profiles exhibited by the hydrodynamic variables in
  these layers are non-monotonic, in contrast with the smooth
  monotonic profiles that have been typically discussed in the
  literature. Thereby, we will refer to them as {\it anomalous} or
  {\it AR} shear layers.

  As mentioned above, jet shear layers, represent natural sites for
  particle acceleration, providing high-energy cosmic rays and
  influencing the dynamics of relativistic jets in extragalactic radio
  sources by forming cosmic-ray cocoons \citep{ost00}. The efficiency
  of the acceleration process in these turbulent shear layers depends
  on the particle mean free path and on the velocity
  structure. \citeauthor{SO02} (2002; SO02 hereafter) performed a
  through study of both the acceleration processes acting at jet
  boundaries and the resulting observational effects.  They former
  authors considered the possibility that turbulent standard boundary
  layers of kiloparsec scale jets may substantially contribute to the
  radiative output of the jet. In the present work we will closely
  follow SO02 working hypothesis, however we will replace the
  monotonic shear layer kinematic structure they assumed by an
  anomalous one as suggested by \cite{AR06}. The aim being to show
  whether AR-shear layers imprint any distinctive feature in the
  radiation produced by the jet as compared with standard boundary
  layers. We show in this paper that, indeed, the radiative output of
  jets bound by anomalous shear layers significantly differs from that
  of ordinary boundary layers. Thus observations may confirm or rule
  out the existence of such anomalous shear layers in kiloparsec scale
  jets and, consistently, this may be used to constrain hardly known
  physical parameters in these regions.

  The plan of the paper is to first show [\S~\ref{sec:RMHD_boost}] the
  influence of magnetic fields in the profiles of the physical
  variables in anomalous boundary layers. In this way we extend the
  results of \cite{AR06} and \cite{Mizuetal08} to the adequate
  parameter range.  In Sect.~\ref{sec:model} we summarize the basic
  model developed by SO02 and adapt it to account for the kinematic
  differences when applied to anomalous shear layers whose profiles
  will be discussed in
  Sect.~\ref{sec:RMHD_boost}. Section~\ref{sec:SED} provides the
  spectral energy distribution obtained for the different models of
  kinematic shear layers shown in the previous section. We discuss our
  results in Sect.~\ref{sec:discussion} and sum them up in
  Sect.~\ref{sec:summary}.

  \section{Magnetohydrodynamic boosting in relativistic jet boundary
    layers}
\label{sec:RMHD_boost}

The dynamics of a relativistic magnetohydrodynamic jet in an external
medium can be treated as the motion of two fluids, one of which (the
jet) is much hotter and at higher (or equal) pressure than the other
one (the external medium), and is moving with a large tangential
velocity with respect to the cold, slowly moving fluid. Furthermore,
the jet can be magnetized while the external medium magnetic field is
much smaller than that of the jet. \cite{AR06} found that if the
specific enthalpy of the jet is sufficiently large, a net conversion
of internal to kinetic energy can be produced through a purely
relativistic channel. The reason for its exclusively relativistic
character is that, the evolution of contact or tangential
discontinuities (like, e.g., the one that separates laterally the jet
from the external medium), is governed by a genuine coupling between
the specific enthalpy and the Lorentz factor at both sides of the
discontinuity which does not exist in Newtonian
(magneto-)hydrodynamics. Such mechanism yields a substantial
hydrodynamic boost along the boundary layer between the jet and the
external medium. Basically, the boost along the jet lateral boundaries
is due to the work done by the external medium on an overpressured,
relativistic jet. \cite{Mizuetal08} extended the validity of the
results of \cite{AR06} to the case in which the jet is magnetized and
conclude that the presence of a jet magnetic field (either poloidal or
toroidal) enlarges the boost that might be obtained by purely
hydrodynamic means.  The magnetohydrodynamic boost is larger for jets
carrying toroidal than poloidal magnetic fields.

The present analysis is aimed to show what is the qualitative
variation of the Lorentz factor and of the magnetic field across a
jet boundary layer. Therefore, following \cite{AR06}, we model the
interaction between the jet and the external medium as a one
dimensional Riemann problem in Cartesian coordinates. Furthermore, we
restrict to the case in which the magnetic field is perpendicular to
the flow speed in the jet. As demonstrated by \cite{AR06} and
\cite{Mizuetal08} this simple model retains the basic features of the
phenomenon. Certainly, a more detailed model would require three
dimensional relativistic magnetohydrodynamic simulations. 

We solve exactly the Riemann problem set by the jump in the conditions
between the jet and the external medium assuming that the magnetic
field is perpendicular to the velocity field. For this purpose we use
the method devised by \cite{Romeroetal05}. The two uniform ``left''
(jet) and ``right'' (external medium) states which set the proposed
Riemann problems possess different and discontinuous
magnetohydrodynamical properties: the rest-mass density $\rho$, the
total pressure $p=p_{\rm mag}+p_{\rm gas}$ ($p_{\rm mag}$ and $p_{\rm
  gas}$ being the magnetic and the gas pressure, respectively), the
components of the velocity normal $v^{\rm n}$ and tangential $v^{\rm
  t}$ to the initial discontinuity, and the magnetization ${\tilde
  \beta}:=p_{\rm mag}/p_{\rm gas}$. In the following, we will use for
the quantities in the jet and in the external medium the subscripts
``L'' and ``R'', respectively.

\begin{figure}
\plotone{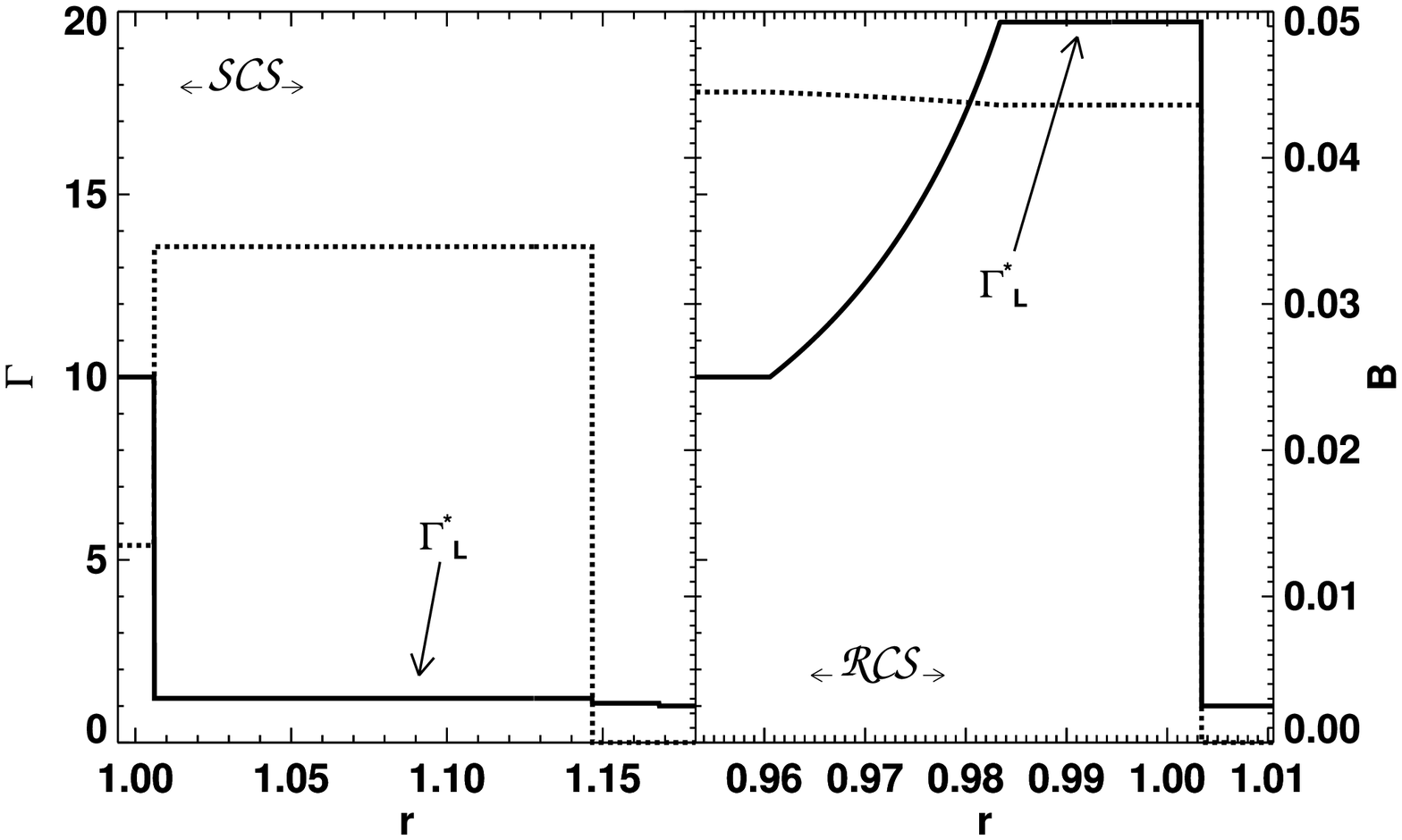}
\caption{{\it Left panel:} Profiles of the Lorentz factor (solid line)
  and of the laboratory frame magnetic field (dashed line) resulting
  from a prototypical Riemann problem yielding a $_{\leftarrow}{\cal
    SCS}_{\rightarrow}$ pattern. The initial left and right states are
  $(p_{_{\rm L}},\rho_{L},v^n_{_{\rm L}},\Gamma_{_{\rm L}},{\tilde \beta}_{\rm
    L})=(10^{-5},10^{-7},0.9,10,0.1)$ and $(p_{\rm
    R},\rho_{R},v^n_{\rm R},\Gamma_{\rm R},{\tilde \beta}_{\rm
    R})=(10^{-6},10^{-2},0,1,0.1)$, respectively. Note that no boost
  is produced to the left of the contact discontinuity located at
  $r\simeq 1.147$. {\it Right panel:} Same as the left panel, but for
  a prototype $_{\leftarrow}{\cal RCS}_{\rightarrow}$ pattern. The
  initial left and right states are $(p_{_{\rm L}},\rho_{L},v^n_{_{\rm
      L}},\Gamma_{_{\rm L}},{\tilde \beta}_{\rm L})=(10^{-5},10^{-7},0,10,100)$
  and $(p_{\rm R},\rho_{R},v^n_{\rm R},\Gamma_{\rm R},{\tilde \beta}_{\rm
    R})=(10^{-6},10^{-2},0,1,0.1)$, respectively. Note the boost (i.e., $\Gamma^*_{_{\rm L}} > \Gamma_{_{\rm L}}$)
  reached by the rarefied jet to the left of the contact discontinuity
  at $r\simeq 1.003$. In both cases the initial discontinuity is set
  at $r=1$ and the solution is shown after an arbitrary amount of time
  (we recall that the solution of the Riemann problem is
  self-similar).
  \label{fig:sl_structure_magnetic}}
\end{figure}

Like in the purely hydrodynamic case, depending on the conditions of
the left and right states two qualitatively and quantitatively
different solutions develop (Fig.~\ref{fig:sl_structure_magnetic}). On
the one hand, a pattern of waves can be formed, which we indicate as
${\cal _{\leftarrow}\!S C S_{\!\rightarrow}}$ (Fig.\@
\ref{fig:sl_structure_magnetic} left panel), where ${\cal
  _{\leftarrow}\!S}$ refers to the shock propagating towards the left
sweeping up the jet, ${\cal S_{\!\rightarrow}}$ to the shock moving
towards the right crossing the external medium and ${\cal C}$ to the
contact discontinuity between the two. On the other hand, the pattern
can change for sufficiently large velocities parallel to the jet axis
(tangential) and in this case the inward-moving shock is replaced by a
rarefaction wave thus producing a ${\cal _{\leftarrow}\!R C
  S_{\!\rightarrow}}$ pattern (Fig.~\ref{fig:sl_structure_magnetic}
right panel). Once again, we point out that the change in the wave
pattern does only happen in relativistic (magneto-)hydrodynamics. Only
in the relativistic regime the evolution of a Riemann problem depends
upon the components of the velocity parallel to the initial
discontinuity or, equivalently, upon the Lorentz factor at both sides
of the initial discontinuity. Furthermore, only in the relativistic
regime there is a coupling between the momentum and energy fluxes that
depends upon the specific enthalpy of initial left and right states.

We notice that the variation of the magnetization in the intermediate
left state is less than a few percent when a ${\cal _{\leftarrow}\!R C
  S_{\!\rightarrow}}$ solution forms (see
Fig.~\ref{fig:sl_structure_magnetic} right panel). In case a ${\cal
  _{\leftarrow}\!S C S_{\!\rightarrow}}$ pattern develops, the
magnetization in the shocked left intermediate state ${\tilde \beta}^*_{_{\rm
    L}}$ grows due to the shock compression, but, in any case,
${\tilde \beta}^*_{_{\rm L}} \simlt (2 - 3)\times{\tilde \beta}_{_{\rm L}}$.

Although \cite{Mizuetal08} have already explored the influence of a
dynamically relevant magnetic field in the jet on the aforementioned
boost, they restricted their study to only four cases and to extremely
hot jets ($p_{_{\rm gas,L}}/\rho_{_{\rm L}} \ge 10^5$), which is
appropriate for gamma-ray burst jets (see, e.g.,
\citealt{alo00b,Alo05,AO07,Birkletal07}). Here we consider a parameter
space more adequate for kiloparsec scale jets. More precisely, we
consider {\it warm} or {\it cold} jets with a ratio $p_{\rm
  gas,L}/\rho_{_{\rm L}} \le 10^2$ and a magnetization ${\tilde
  \beta}_{_{\rm L}}$, which is varied between $0.1$ (slightly
magnetized jet) and $10^4$ (Poynting flux dominated jet)\footnote{In
  \cite{Mizuetal08} a maximum value of ${\tilde \beta}_{_{\rm L}}=1.8$
  is considered in their model MHDB.}.  The left state will have
comparable but larger total pressure than the external medium
$p_{_{_{\rm L}}} = 10 p_{_{\rm R}}$, will be under dense $\rho_{_{\rm
    L}} / \rho_{_{\rm R}} = 10^{-5}$, with a fixed bulk Lorentz factor
$\Gamma_{_{\rm L}} \equiv [1 - (v^t_{_{\rm L}})^2 - (v^n_{_{\rm
    L}})^2]^{-1/2}=10$. We choose to fix the external medium (right
state) to be cold ($p_{_{\rm R}}/\rho_{_{\rm R}} \simeq 10^{-4}$),
non-magnetized and non-moving. Precisely, we take for the external
medium $p_{_{\rm R}}=10^{-6}/\rho_{_{\rm ext}} c^2$, $\rho_{_{\rm R}}
= 10^{-2}\rho_{_{\rm ext}}$, $v^n_{_{\rm R}} = v^t_{_{\rm R}} =
{\tilde \beta}_{_{\rm L}} = 0$, where $\rho_{_{\rm ext}}$ is a
normalization constant which allows us to be scale-free. We point out
that with this parametrization we provide physically plausible
conditions for both the jet and the external medium.  For convenience,
hereafter we will assume $c = \rho_{_{\rm ext}} = 1$, unless stated
otherwise.

\begin{figure}
\epsscale{0.82}
\plotone{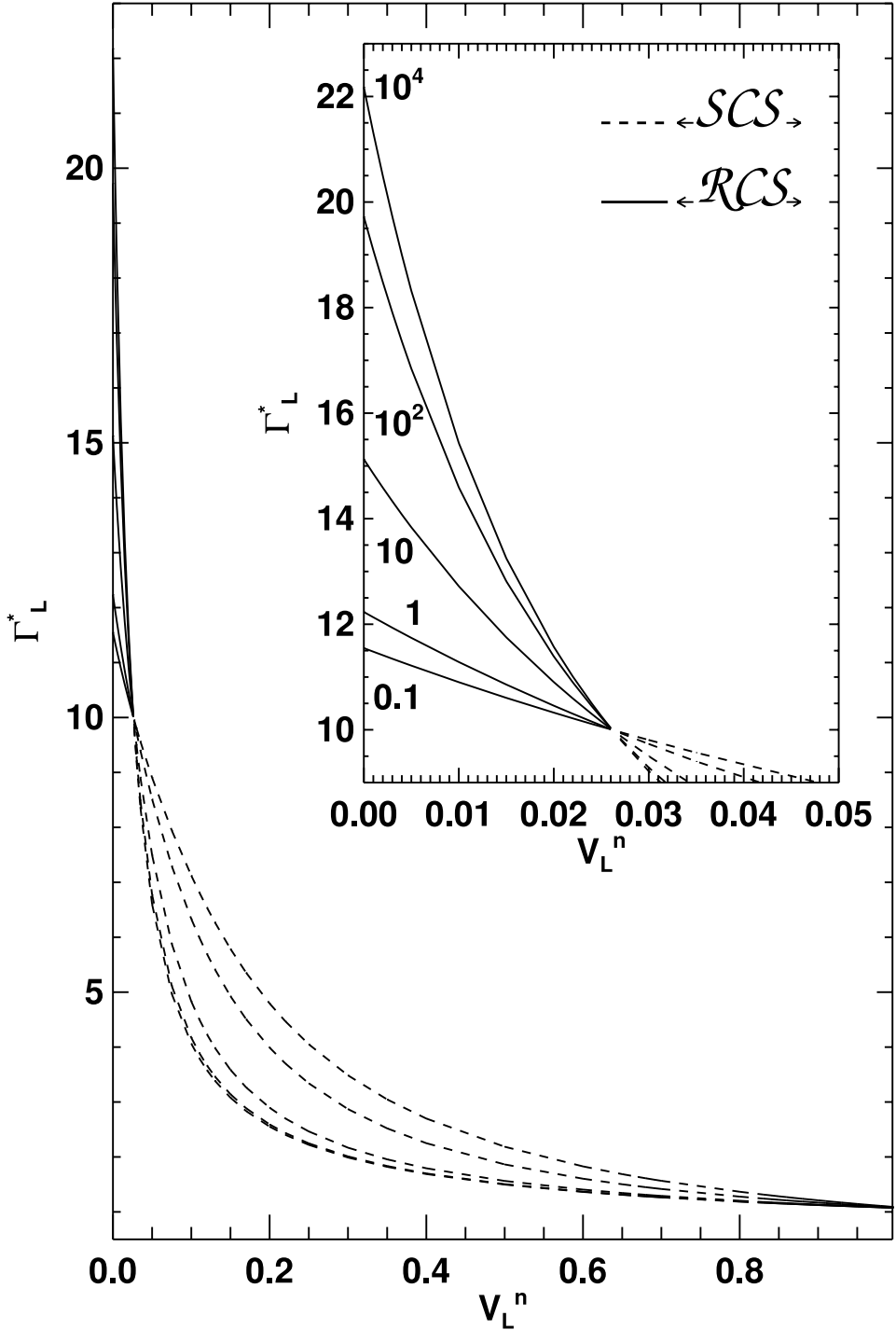}
  \caption{Lorentz factor reached at the uniform state left to the
    contact discontinuity $\Gamma^*_{_{\rm L}}$ as a function of the
    normal velocity in the left state $v^n_{_{\rm L}}$. The right
    state is held fixed and is given by $(p_{_{\rm R}}, \rho_{_{\rm
        R}}, v^n_{_{\rm R}}, \Gamma_{_{\rm R}},{\tilde \beta}_{_{\rm R}}) =
    (10^{-6}, 10^{-2}, 0, 1, 0.1)$. The left state has fixed values in
    the total pressure, the rest-mass density and the Lorentz Factor
    $(p_{_{\rm L}}, \rho_{_{\rm L}}, \Gamma_{_{\rm L}})=(10^{-5},
    10^{-7}, 10)$, while the magnetization ${\tilde \beta}_{_{\rm L}}$ is
    varied for each curve as indicated by the different labels (see
    inset). The solid lines refer to the $_{\leftarrow}{\cal
      RCS}_{\rightarrow}$ pattern, and the dashed lines refer to a
    $_{\leftarrow}{\cal SCS}_{\rightarrow}$ pattern. The inset shows a
    zoom of the solutions for very small values of $v^n_{_{\rm L}}$,
    which are the most expected ones for {\it standard} jet flows.
    \label{fig:betal_dep}}
\end{figure}

We explore the resulting Riemann solution as we vary the component of
the jet velocity normal to the shear layer\footnote{For a relativistic
  jet it is expected that such a component is $v^n_{_{\rm L}} \simeq 0
  \ll v^t_{_{\rm L}} \sim 1$.} and the jet magnetization ${\tilde
  \beta}_{_{\rm L}}$. We point out that models with ${\tilde \beta}
\ge 100$ are actually rather cold, since $p_{\rm gas,L}/\rho_{_{\rm
    L}} \ll 1$ (e.g., for ${\tilde \beta} \ge 10^4$, $p_{\rm
  gas,L}/\rho_{_{\rm L}} \simeq 10^{-2}$). According to the most
broadly accepted view, even MHD-generated jets may become matter
dominated at kiloparsec scale (e.g., \citealp{BL94,FO04}). However,
there are others who advocate jet models which are basically
electromagnetic entities at such large scales (e.g.,
\citealp{Blandford02,Blandford03}). In view of these two possible
extremes, we have covered the ${\tilde \beta}$-parameter space with
jet models which are both matter dominated and Poyinting-flux
dominated at kiloparsec scale. In Fig.~\ref{fig:betal_dep} we show the
value of the Lorentz factor reached at the state left to the contact
discontinuity $\Gamma^*_{_{\rm L}}$. This intermediate state is made
out of boosted or deboosted jet matter depending respectively on
whether a ${\cal _{\leftarrow}\!R C S_{\!\rightarrow}}$ or a ${\cal
  _{\leftarrow}\!S C S_{\!\rightarrow}}$ solution develops.

Figure~\ref{fig:betal_dep} illustrates how a magnetohydrodynamic
boost, where $\Gamma^*_{_{\rm L}} > \Gamma_{_{\rm L}}$, exists below
a critical value of the normal velocity ($v^n_{_{\rm L}}\simeq 0.026$
in the case considered here) whereby a ${\cal _{\leftarrow}\!R C
  S_{\!\rightarrow}}$ solution develops. The critical value of
$v^n_{_{\rm L}}$ is independent of the jet magnetization (note the
crossing of all the solutions at the point where $\Gamma^*_{_{\rm L}}
= \Gamma_{_{\rm L}}=10$). 

The magnetohydrodynamic boost is larger if the magnetization of the
jet is increased: while for a poorly magnetized jet (${\tilde
  \beta}_{_{\rm L}}=0.1$) the boost may increase the bulk Lorentz
factor of the layer by $\sim 10\% - 50\%$, for a strongly magnetized
jet (${\tilde \beta}_{_{\rm L}}\ge 10^2$) the Lorentz factor increase
can be $\ge 100\%$. Increasing the magnetization beyond ${\tilde
  \beta}_{_{\rm L}}\ge 10^4$ does not produce larger boosts (the
effect saturates). Below ${\tilde \beta}_{_{\rm L}} < 0.1$ the results
are almost indistinguishable from the case ${\tilde \beta}_{_{\rm
    L}}=0.1$ and we have decided not to include more lines in the
Fig.~\ref{fig:betal_dep} for the sake of readability.

The bottom line of this parametric study is that for conditions
realizable in kiloparsec scale jets, one may find an increase of the
Lorentz factor in the transition layer between the jet and the
external medium ranging from $(1.5 - 2) \times\Gamma_{_{\rm L}}$,
while the magnetization in that layer is basically the same as in the
jet if a ${\cal _{\leftarrow}\!R C S_{\!\rightarrow}}$ pattern
occurs. Furthermore, since only the strength of the
magnetohydrodynamic boost, but not its existence, depends on the
topology of the field (see \citealt{Mizuetal08}), we point out that it
also happens in the presence of randomly oriented magnetic fields.

\section{The physical model} 
\label{sec:model}

\begin{figure}
\plotone{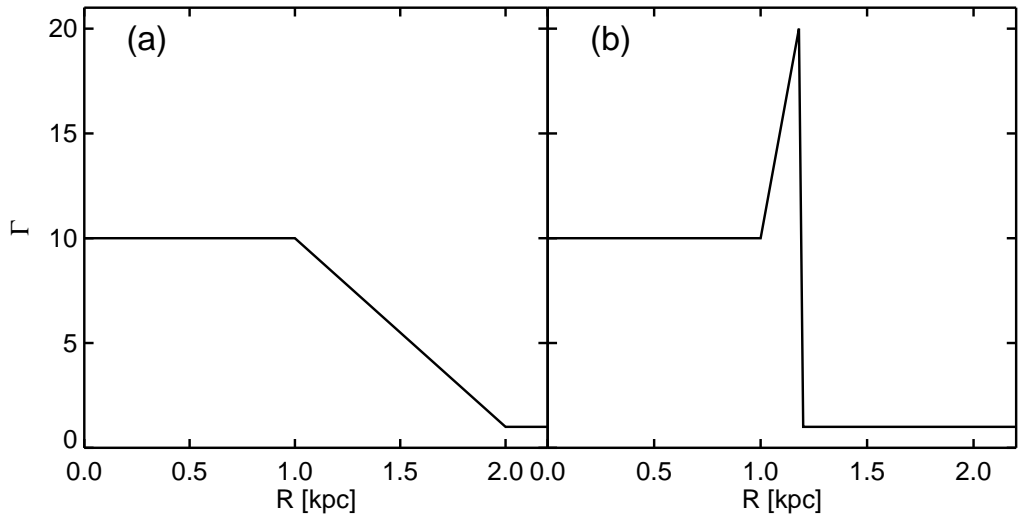}
\caption{Radial profile of the Lorentz factor across the jet showing
  the kinematic difference in the structure of an standard, monotonic
  shear layer (panel a) and an anomalous shear layer structure (panel
  b).
\label{fig:sl_structure}}
\end{figure}

Our model follows very closely that devised by SO02. We assume that
non-thermal particles can be accelerated at the boundary layers of
relativistic jets (e.g., \citealp{ost90,ost98,ost00,RD04}), thereby
producing high-energy cosmic rays. As mentioned in
Sect.~\ref{sec:intro}, the main goal of this paper is to replace the
prototype shear layer assumed in SO02 by an anomalous shear layer of
the type suggested by \cite{AR06} and whose basic features have been
outlined in Sect.~\ref{sec:RMHD_boost}. At the most basic level, this
means to replace the kinematic shear layer structure assumed by SO02,
namely, a jet sheath where the bulk Lorentz factor ($\Gamma$)
decreases monotonically (Fig.~\ref{fig:sl_structure}a), by a boundary
layer where the Lorentz factor develops a spike in which it is larger
than at the jet core (Fig.~\ref{fig:sl_structure}b). We shall assume
that the magnetic field is uniform in the whole shear layer, which is
compatible with the results of our previous section and a reasonable
assumption in our simple model.

The working hypothesis made by SO02 are also valid in our case since,
the only difference between monotonic and anomalous layers is simply
kinematic (see Sect.~\ref{sec:kinematic}). Thus, following SO02, the
maximum electron Lorentz factor which can be obtained by the
combination of acceleration of particles along the shear layer and the
radiative cooling of such particles is
\begin{equation}
\gamma_{\rm eq} \approx 4 \cdot 10^{8} \, V_{8} \, B_{\mu G}^{-1 / 2} , 
\label{eq:gamma_eq}
\end{equation}
where $B_{\mu G}$ is the magnetic field in microgauss and $V_8\equiv
V_{\rm A}/10^8\,$cm\,s$^{-1}$ is the Alfv\'en speed $V_{\rm A}$ in
units of $10^8\,$cm\,s$^{-1}$. For kiloparsec scale jets, typical
values of these two parameters are $V_8\simeq 1$ and $B_{\mu G}\sim
10$, which yields $\gamma_{\rm eq} \simeq 10^8$.

We assume that the resulting spectral energy distribution of the
electrons accelerated at the layer consists of a low energy power law
$n_{\rm e}(\gamma) \propto \gamma^{-\sigma}$, with $\sigma = 2$,
finished with a high-energy component modeled as a nearly mono
energetic peak at $\gamma = \gamma_{eq}$ due to the pile-up of
accelerated particles caused by losses (see Fig.~1 of SO02). In terms
of the Dirac delta function, $\delta$, at energies above the injection
energy, $\gamma > \gamma_0$, the considered quasy-stationary, averaged
spectrum of electrons can be cast in the form
\begin{equation}
n_{\rm e}( \gamma) = a \, \gamma^{- \sigma} \, \exp( - \gamma / \gamma_{\rm eq}) + b \, \delta( \gamma - \gamma_{\rm eq}) , 
\label{eq:n(gamma)}
\end{equation}
where $a$ and $b$ are normalization parameters whose values ($a\simeq
10^{-7}\,$cm$^{-3}$, $b \simeq 10^{-14}\,$cm$^{-3}$) result from
assuming equipartition between the magnetic field energy density
$u_{_{\rm B}} \equiv B^2We / 8 \, \pi$ and the energy densities of both
electron spectral components:
\begin{equation}
\int_{\gamma_0}^\infty (\gamma \, mc^2) \, a \, \gamma^{-2} \, \exp(
- \gamma / \gamma_{\rm eq})  \, d \gamma = \int_{\gamma_0}^\infty  (\gamma
\, mc^2) \, b \, \delta( \gamma - \gamma_{\rm eq}) \, d \gamma = {1 \over
  2} \, u_{_{\rm B}} . 
\label{eq:normalization}
\end{equation}

We point out that in SO02, the power-law part of the electron spectrum
was also endowed with an exponential cut-off instead of with a
Heaviside $\Theta$ function in spite of their Eqs.~6 and 7. The
$\Theta$ function was used in their work only to compute analytic
estimates, not for the numerical integrations necessary to calculate
their spectral energy distributions \citep{Stawarz}.

Assuming equipartition between the particle and magnetic field energy
in relativistic sheared jets has been proved to be a rather solid
theoretical assumption (see, e.g., \citealt{Urpin06}). However, in
some cases the values of $a$ and $b$ derived from
Eq.~\ref{eq:normalization} and based on equipartition arguments may
yield only lower bounds of the true values. \cite{KS05} argue that
powerful jets in quasars and FR II objects can be far from the minimum
energy condition, and the field strength very likely exceeds the
equipartition value. On the other hand, upper limits to the inverse
Compton radiation of the jet in M\,87 imposed by HESS and HEGRA
Cerenkov Telescopes also indicate that the magnetic field cannot be
weaker than the equipartition value \citep{Stawarzetal05}.

\subsection{Radiative processes in boundary layers of kpc-scale jets}
\label{sec:radiation}
%
%

The relevant radiative processes taking place at the boundary layer of
a relativistic, kpc-scale jet are, on the one hand, the synchrotron
(syn) emission of the spectral family represented by
Eq.~\ref{eq:n(gamma)} and, on the other hand, their inverse Compton
(IC) cooling due to the interaction with the previously produced
synchrotron photons (synchrotron self Compton -SSC-) or with seed
photons of the cosmic microwave background (external Compton
-EC-). These radiative processes are the dominant for large scale
relativistic ($\Gamma>2$) jets at distances from the galactic nucleus
$z> 10\,$kpc.

Following SO02, we assume that the magnetic field is randomly oriented
in the shear layer\footnote{Note that anomalous layers can also form
  if the magnetic field is randomly oriented, see
  Sect.~\ref{sec:RMHD_boost}.}. Additionally, we neglect synchrotron
self-absorption and, therefore, we limit the computation of the
resulting spectra to frequencies $\nu\geq 10^{10}\,$Hz.

Restricted to the case of an isotropic electron distribution
$n_{\rm e}(\gamma)$ the synchrotron emissivity averaged over a randomly
oriented magnetic field $B$ can be computed from
\begin{equation}
  j_{\rm syn}(\nu) = { \sqrt{3} e^{3} B \over mc^{2}} \int R \left( {\nu
      \over c_{1} \gamma^{2}} \right) \, {n_{e}(\gamma) \over 4 \pi}
  \, d\gamma , 
\label{eq:j_syn}
\end{equation}
where $c_1 = 3 e B \, / \, 4 \pi m c$ and $R(x)$ is a combination of
the modified second order Bessel functions \citep{cru86} as
\begin{equation}
  R(x) = {x^2 \over 2} \, K_{4/3} \left({x \over 2} \right) \, K_{1/3}
  \left({x \over 2} \right) - 0.3 \, {x^3 \over 2} \left[ K_{4/3}^2
    \left({x \over 2} \right) - K_{1/3}^2 \left({x \over 2} \right)
  \right] . 
  \label{eq:R(x)}
\end{equation}

%
%
The IC photon emissivity in the source frame%
\footnote{In our case the source frame is a cylindrical layer of the
  boundary region where $\Gamma$ is assumed to be uniform.},
$\dot{n}_{\rm ic} (\epsilon, \Omega)$, at a certain photon energy
$\epsilon$ (in units of the electron rest mass, $m_{\rm e}c^2$,
$\epsilon \equiv h \nu / m_{\rm e}c^2$) and scattering direction
$\Omega$, is given by
\begin{equation}
  \dot{n}_{\rm ic}(\epsilon,\Omega) = c \int d\epsilon_i \oint d\Omega_i
  \int d\gamma \oint d\Omega_{\rm e} \, (1- \beta \, \cos \psi) \, \sigma \,
  n_i(\epsilon_i,\Omega_i) \, n_{\rm e}(\gamma,\Omega_{\rm e}) , 
\label{eq:n_ic}
\end{equation}
where $n_i (\epsilon_i, \Omega_i)$ is the seed photon number density
as a function of the incident photon energy $\epsilon_i$ and the
photon direction $\Omega_i$,
$n_{\rm e} (\epsilon_{\rm e}, \Omega_{\rm e})$ is the electron energy
distribution, $\psi$ is the angle between the electron and the
incident photon directions, $\gamma=(1-\beta)^{1/2}$ is the electron
Lorentz factor corresponding to a velocity $v_{\rm e}=\beta c$ and
$\Omega_{\rm e}=(\cos^{-1}{\mu_{\rm e}}, \phi_{\rm e})$ its direction
in the plasma source frame (e.g., \citealt{der95}). The rest-frame
emissivity in CGS units is $j_{\rm ic}(\nu,\Omega) = h \, \epsilon \,
\dot{n}_{\rm ic} (\epsilon, \Omega)$.

In the calculation of the emissivity due to IC process (SSC or EC), we
limit ourselves to the Thompson regime, in which the scattering cross
section $\sigma \equiv \sigma_{_{\rm T}}=6.65\times 10^{-25}\,$cm$^2$ is
independent of the seed photon energy, which is an adequate
approximation in the regime $\gamma \epsilon_i < 1$. 

Following \cite{der95}, the scattered photons will be beamed along the
direction of the electron motion (i.e. $\Omega\approx\Omega_{\rm e}$)
and will emerge after scattering with an average final energy
$\epsilon\approx\frac{4}{3}\gamma^2\epsilon_{\rm syn}$. We assume as
SO02 that the synchrotron radiation and the electron distribution are
isotropic and, therefore, $n_i(\epsilon_i,\Omega_i) = n_{\rm
  syn}(\epsilon_{\rm syn})/4\pi$ and $n_{\rm e}(\gamma,\Omega_{\rm e})
= n_{\rm e}(\gamma)/4\pi$. Then, in Eq.~\ref{eq:n_ic} one can use
$\sigma = \sigma_{_{\rm T}} \delta(\Omega-\Omega_{\rm e}) \delta
\left(\epsilon-(4/3)\gamma^2\epsilon_{\rm syn}\right)$ in order to
obtain the following expression of $\dot{n}_{_{\rm
    SSC}}(\epsilon,\Omega)$:
\begin{equation}
  \dot{n}_{_{\rm SSC}}(\epsilon,\Omega) =
  \frac{c\sigma_{_{\rm T}}}{4\pi}\int_{0}^{\infty} {d\epsilon_{\rm
      syn}}\int_{\gamma_0}^{\infty} {d\gamma}n_{\rm syn}(\epsilon_{\rm
    syn}) n_{\rm
    e}(\gamma)\delta\left(\epsilon-\frac{4}{3}\gamma^2\epsilon_{\rm syn}\right).
\label{eq:n_ssc}
\end{equation}
Using Eqs.~\ref{eq:n(gamma)},~\ref{eq:n_ssc}, the relation $j_{_{\rm
    SSC}}(\nu,\Omega) = h\epsilon \dot{n}_{_{\rm
    SSC}}(\epsilon,\Omega)$ and the optically thin synchrotron
intensity $I_{\rm syn}(\nu_{\rm syn})\equiv j_{\rm
  syn}l=(ch/4\pi)\epsilon_{\rm syn}n_{\rm syn}(\epsilon_{\rm syn})$,
where $l$ is the emitting region linear size, the integral form of the
synchrotron self-Compton emissivity $j_{_{\rm SSC}}(\nu)$ is written
as
\begin{equation}
  j_{_{\rm SSC}}(\nu)=\sigma_{_{\rm T}} \int_{\gamma_{_{\rm low}}}^{\infty} {j_{\rm syn}\left(\frac{3\nu}{4\gamma^2}\right)l\frac{4\gamma^2}{3}n_{\rm e}(\gamma)d\gamma},
\label{eq:j_ssc}
\end{equation}
\noindent
where $\gamma_{_{\rm low}}={\rm max}(\gamma_0, \frac{3}{4}\epsilon)$ is
the lower limit imposed by the restriction $\gamma \epsilon_{\rm
  syn}<1$, which guaranties that we stay within the Thompson regime.

%
%
The CMB radiation is strongly anisotropic in the source frame which is
moving at a bulk Lorentz factor $\Gamma$, providing an external source
of seed photons which move opposite to the jet direction, i.e., $\psi
\simeq \pi$. Thereby, $(1-\beta \cos \psi)=(1+\mu_{\rm e})$, where
$\Omega_{\rm e}=(\cos^{-1}{\mu_{\rm e}}, \phi_{\rm e})$ is the
direction of the colliding electron in the comoving frame of the shear
layer, similarly to what we made for the case of SSC radiation. In the
Thomson regime $\sigma=\sigma_{_{\rm T}}\delta(\Omega-\Omega_{\rm
  e})\delta \left[ \epsilon-\gamma^2\epsilon_{\rm cmb}(1+\mu_{\rm
    e})\right]$, since the scattered photon energy is
$\epsilon=\gamma^2\epsilon_{\rm cmb}(1+\mu_{\rm e})$, where
$\epsilon_{\rm cmb}$ is the energy of the incident CMB photon in units
of the electron rest-mass.  Hence, Eq.~\ref{eq:n_ic} can be cast as
\begin{equation}
  \dot{n}_{_{\rm EC}}(\epsilon,\Omega)=\frac{c\sigma_{_{\rm T}}}{4\pi}(1+\mu)\int_{0}^{\infty}
  {d\epsilon_{\rm cmb}}\int_{\gamma_0}^{\infty}
  {d\gamma}n_{\rm cmb}(\epsilon_{\rm cmb})n_{\rm e}(\gamma)\delta\left[\epsilon-\gamma^2\epsilon_{\rm cmb}(1+\mu)\right]\  .
\label{eq:n_ec}
\end{equation}

In order to compute the EC emissivity produced by the interaction of
CMB photons with the power-law part of the electron distribution, we
model the CMB field as a monochromatic radiation with photon energies
equal to the average value of a thermal black body spectrum
\citep{BG70} but blue-shifted by a factor $\Gamma$, i.e.,
$\epsilon_{\rm cmb} = \Gamma \langle \epsilon_{\rm cmb}^{\ast} \rangle \equiv 2.7
\times \Gamma \, k \, T^{\ast} / \me c^2$, and a number density
$n_{\rm cmb} = \Gamma \, u_{\rm cmb}^{\ast} / (\langle \epsilon_{\rm
  cmb}^{\ast} \rangle \me c^2)$. The value that we adopt for the energy
density of the CMB radiation in the rest frame of the CMB is $u_{\rm
  cmb}^{\ast}=4\cdot 10^{-13}\,$erg\,cm$^{-3}$. With these
approximations, the EC emissivity associated to the power-law part of
the electron distribution can be computed analytically 
\begin{eqnarray}
  j_{_{{\rm EC},1}}(\nu) &=&
        \displaystyle{\frac{c \sigma_{_{\rm T}} a}{8\pi} u_{\rm cmb}^{\ast}
          \left(\frac{h}{2.7kT^\ast} \right)^{1/2}
          [(1+\mu)\Gamma]^{3/2}
          \exp{\left[-\left(\frac{\epsilon}{\Gamma (1+\mu) 
           \langle \epsilon_{\rm cmb}^\ast \rangle \gamma_{\rm eq}^2} \right)^{1/2}
           \right]}\nu^{-1/2}\:\:} \nonumber \\ 
& &\mbox{if } 
        \gamma_0^2 \le \displaystyle{
          \frac{h\nu}{2.7kT^\ast\Gamma(1+\mu)}}\, . 
\label{eq:j_ec_1}
\end{eqnarray}

For the monoenergetic electron peak component, we use the exact black
body spectrum transformed to the source frame. In this case, the
expression of $n_{\rm cmb}(\epsilon_{\rm cmb})$ in the boundary layer
comoving frame is (see Eq.~2.58 of \citealp{BG70})
\begin{equation}
  n_{\rm cmb}(\epsilon_{\rm cmb})=8\pi \me 
  \left(
    \frac{\me c^2\epsilon_{\rm cmb}}{h}
  \right)^2
  \left[ \exp{ \left( \displaystyle\frac{\Gamma \me c^2\epsilon_{\rm cmb}}{kT^*}
      \right) } - 1
  \right]^{-1}\, ,
\label{eq:n_cmb}
\end{equation}
where $k$ is the Boltzmann constant. Substituting Eq.~\ref{eq:n_cmb}
and $n_{\rm e}(\gamma) = b\delta(\gamma - \gamma_{\rm eq})$ into
Eq.\ref{eq:n_ec}, taking into account that $j_{_{\rm
    EC}}(\nu,\Omega)=h\epsilon \dot{n}_{_{\rm EC}}(\epsilon,\Omega)$
and the restriction imposed by the Thompson regime ($\epsilon_{\rm
  cmb} \gamma_{\rm eq} < 1$), one obtains
\begin{eqnarray}
  j_{_{{\rm EC},2}}(\nu)&=&
  \displaystyle{\frac{2 h \sigma_{_{\rm T}} b}
    {(1+\mu)^2 c^2 \gamma_{\rm eq}^6} 
    \left(\exp{\left(\frac{\Gamma h \nu}{(1+\mu)kT^\ast}\right)}-1 \right)^{-1}
    \nu^3} \nonumber \\
  & & \mbox{ if } 
  \nu \le \displaystyle{
    \frac{(1+\mu) \gamma_{\rm eq} m_{\rm e} c^2}{h} }\, .
\label{eq:j_ec_2}
\end{eqnarray}

The total EC emissivity is the sum of Eqs.~\ref{eq:j_ec_1} and
\ref{eq:j_ec_2}, i.e.,
\begin{equation}
  j_{_{\rm EC}}(\nu) = j_{_{{\rm EC},1}}(\nu) + j_{_{{\rm EC},2}}(\nu)\, .
\label{eq:j_ec}
\end{equation}

\subsection{Kinematic model of the shear layer}
\label{sec:kinematic}

\begin{figure}
\epsscale{0.45}
\plotone{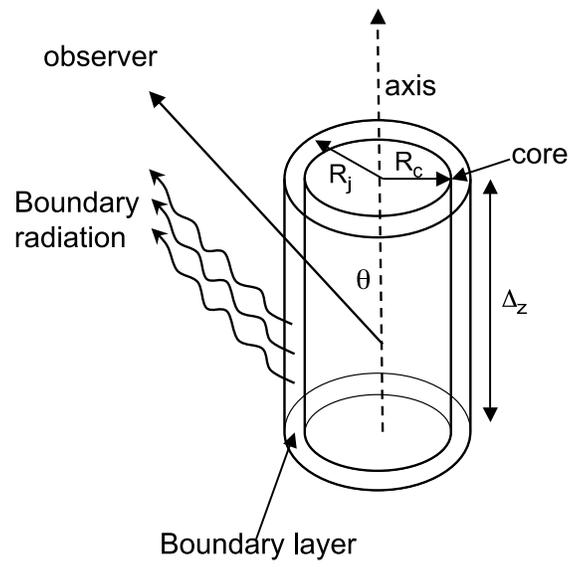}
\caption{Schematic view of the jet model.
  \label{fig:scheme}}
\end{figure}

We model a large scale jet as a uniform cylindrical spine of radius
$R_{\rm c}$ with a Lorentz factor $\Gamma_{\rm c}$ followed by a
transition layer where the Lorentz factor changes linearly with radius
until $r=R_{\rm j}$, where the flow velocity either becomes zero
(similar to the SO02 model) or reaches a maximum from which it
abruptly decreases to zero (Figs.~\ref{fig:sl_structure},
\ref{fig:scheme}). For completeness, we also consider a boundary layer
where the Lorentz factor is uniform. More explicitly
\begin{equation}
  \Gamma_{i}(r) = \left\{ \begin{array}{ll}
      \Gamma_{\rm c} &\mbox{$r \leq R_{\rm c}$} \\
      \Gamma_{{\rm j}, i} + (\Gamma_{\rm c} - \Gamma_{{\rm j}, i}) 
      \frac{R_{\rm j} - r}{R_{\rm j} - R_{\rm c}} &\mbox{$R_{\rm c} < r
        \leq R_{\rm j}$} \\
      1 &\mbox{$r>R_{\rm j}$}
       \end{array} \right.  , 
\label{eq:Gamma}
\end{equation}
where we take $\Gamma_{\rm c}=10$ for our reference models. The
subscript $i=1$ labels the case of a monotonic shear layer
($\Gamma_{{\rm j},1}=1$), $i=2$ represents the case of an anomalous
shear layer ($\Gamma_{{\rm j},2}=15$) and $i=3$ denotes the case of a
jet where the boundary layer is uniform and has the same Lorentz
factor as the core ($\Gamma_{{\rm j},3}=10$). The kinematic
idealization of the case corresponding to an anomalous layer is a
rough prototype of the structure displayed by the Lorentz factor when
a ${\cal _{\leftarrow}\!R C S_{\!\rightarrow}}$ solution develops
(Fig.~\ref{fig:sl_structure_magnetic}). We neglect the variation of
the magnetic field across the boundary layer because it is rather
small (see Sect.~\ref{sec:RMHD_boost}) and we assume that the magnetic
field is uniform and equal to $10\,\mu$G.

In order to properly normalize the emissivity properties of the
different jet models, we choose that all the shear layers transport
the same kinetic power as the monotonically decreasing one with a jet
radius $R_{{\rm j},1} = 2\,$kpc and a density and pressure equal to
the values chosen for the left state of the models shown in
Sect.~\ref{sec:RMHD_boost}. This means that we adjust the external jet
radius of the other two cases to $R_{{\rm j},2} = 1.25\,$kpc and
$R_{{\rm j},3} = 1.38\,$kpc. Since, typically, for extragalactic jets,
$\rho_{_{\rm ext}}\sim 10^{-27}-10^{-24}\,$g\,cm$^{-3}$
\citep{Ferrari98}, the fluxes of mass and of magnetic field in the jet
axial direction are
\begin{eqnarray}
  \Phi &=& 3\cdot 10^{38} \left(\frac{B}{10^{-4}\,{\rm G}}\right) 
    \left(\frac{R_{\rm j}}{1\,{\rm kpc}}\right)^2 \,{\rm G\,cm}^{-2} \\
    {\dot M} &=& 9\cdot10^{25}\left(\frac{R_{\rm c}}{1\,{\rm
          kpc}}\right)^2 \left(\frac{\rho_{_{\rm ext}}}{10^{-24}\,{\rm
          g\,cm}^{-3}}\right)\, {\rm g\,s}^{-1}\, .
\end{eqnarray}

For each cylindrical shell in the shear layer of radius $r$ and moving
with a bulk Lorentz factor $\Gamma(r)$, we compute the local source
frame emissivities for the radiative processes considered in
Sect.~\ref{sec:model} (Eqs.~\ref{eq:j_syn}, \ref{eq:j_ssc} and
\ref{eq:j_ec}). Under the assumption that the radiating electrons are
distributed uniformly in the boundary layer (for a comoving observer),
the emissivity of the sub-layer at the distance $r$ from the jet axis
can be computed as
\begin{equation}
  j^{\ast}_I (\nu^{\ast}, \theta^{\ast}, r) =
  {\cal D}^2(r,\theta^{\ast}) \, j_I\left( \nu = {\nu^{\ast} \over
      {\cal D}(r,\theta^{\ast} )}, \, \mu = { \mu^{\ast} - \beta(r)
      \over 1 - \beta(r) \, \mu^{\ast}} \right) , 
\label{eq:j^ast}
\end{equation} 
where ${\cal D}(r, \theta^{\ast}) \equiv \left[ \Gamma(r) (1 -
  \beta(r) \, \mu^{\ast})\right]^{-1}$ is the Doppler factor
associated to a sub-layer moving with the Lorentz factor $\Gamma(r) =
\left[ 1 - \beta^{2}(r) \right]^{-1/2}$ at an angle $\theta^{\ast}
\equiv \cos^{-1} \mu^{\ast}$ with respect to the line of sight. The
subscript $I =$ syn, SSC or EC specifies the radiative process.

The observed flux density from the boundary layer volume $V^{\ast}$
appropriately modified by the flow-beaming patterns is
\begin{equation}
  S^{\ast} (\nu^{\ast}, \theta^{\ast}) = 
          d^{-2} \int_{R_{\rm c,i}}^{R_{\rm j,i}} dV^{\ast} \, j^{\ast}(\nu^{\ast}, \theta^{\ast},
          r),
\label{eq:S(nu,theta)}
\end{equation}
$d$ being the distance to the observer (we take $d=10\,$Mpc unless
stated otherwise). $dV^{\ast}=2\pi \Delta_{\rm z}r dr$ corresponds to a
volume element of the shear layer with an observed length $\Delta_{\rm
  z}$, which we assume to be $\Delta_{\rm z}=1\,$kpc.  The integrals
involved in the evaluation of Eq.~\ref{eq:S(nu,theta)} are performed
numerically and, in order to compute the SSC contribution to the flux
density, we estimate the SSC emissivity (Eq.~\ref{eq:j_ssc}) by taking
$l=(R_{{\rm j},i} - R_{{\rm c},i})/\sin{\theta^{\ast}}$.

\section{Spectral energy distribution of different models of boundary
  layers for kpc-scale jets}
\label{sec:SED}

\begin{figure}
\plotone{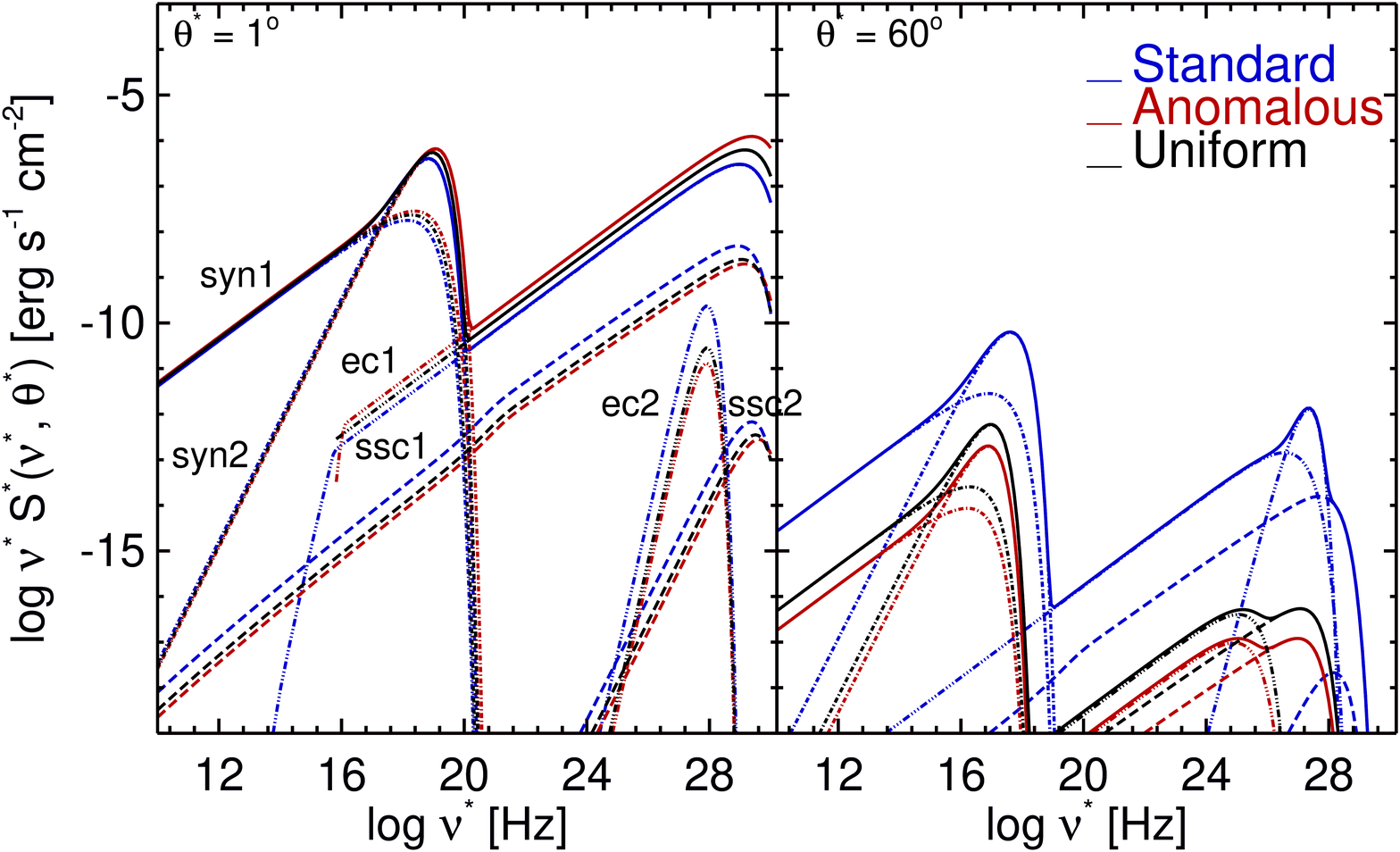}
\caption{Predicted spectral energy distributions of the radiation
  generated in several boundary layers of a kpc-scale jet at
  $\theta^\ast=1^\circ$ and $\theta^\ast=60^\circ$ assuming
  $\gamma_{\rm eq}=10^8$. The blue lines corresponds to the same
  monotonic shear layer profile as in SO02, the black lines to a
  uniform boundary layer and the red lines to an anomalous one (see
  Sect.~\ref{sec:kinematic}). The different contributions to the
  spectrum are labeled above the corresponding lines. The labels 1 and
  2 refer to the components associated to the power-law and to the
  monoenergetic parts of the electron distribution
  (Eq.~\ref{eq:n(gamma)}), respectively.
    \label{fig:SED_AR-SO}}
\end{figure}

Our main goal is to outline the observational differences between the
emission of large-scale jets laterally endowed by different types of
shear layer.  The observed spectral energy distribution of the
radiation emitted by different kinematic boundary layers in a
prototype kpc-scale jet is show in Fig.~\ref{fig:SED_AR-SO} for two
different viewing angles $\theta^\ast = 1^\circ$ and $\theta^\ast =
60^\circ$. We neglect both the absorption of very high energy (VHE)
gamma-rays during the propagation to the observer and any contribution
from a separate non-thermal electron population accelerated at shocks
happening at the jet core. The choice of a small and large viewing
angle is motivated by the fact that for small viewing angles,
$\theta^\ast \simlt 4 \simeq 1/\Gamma_{{\rm j},2}$, the larger Doppler
boosting provided to the emitted radiation by the larger Lorentz
factor in AR layers than in monotonic layers makes the total flux
density of the former case also slightly larger
(Fig.~\ref{fig:SED_AR-SO} left panel). The differences in the total
flux density are smaller in the synchrotron dominated part of the
spectrum than in the EC dominated one.  At larger angles
(Fig.~\ref{fig:SED_AR-SO} right panel), the flux density predicted for
standard layers is 2 to 5 orders of magnitude larger than for uniform
or AR layers due to the Doppler hiding of the later two cases. The
differences are very important in the EC dominated part of the
spectrum (i.e., in the gamma-ray regime).

Both, anomalous and uniform boundary layers generate a rather similar
spectral pattern, particularly at small viewing angles. As we observe
the jet at larger values of $\theta^\ast$ (Fig.~\ref{fig:SED_AR-SO}
right panel) the smallest Lorentz factor of the uniform case provides
a smaller Doppler hiding which results into a larger flux density
than in case of AR layers.

\begin{figure}
\plotone{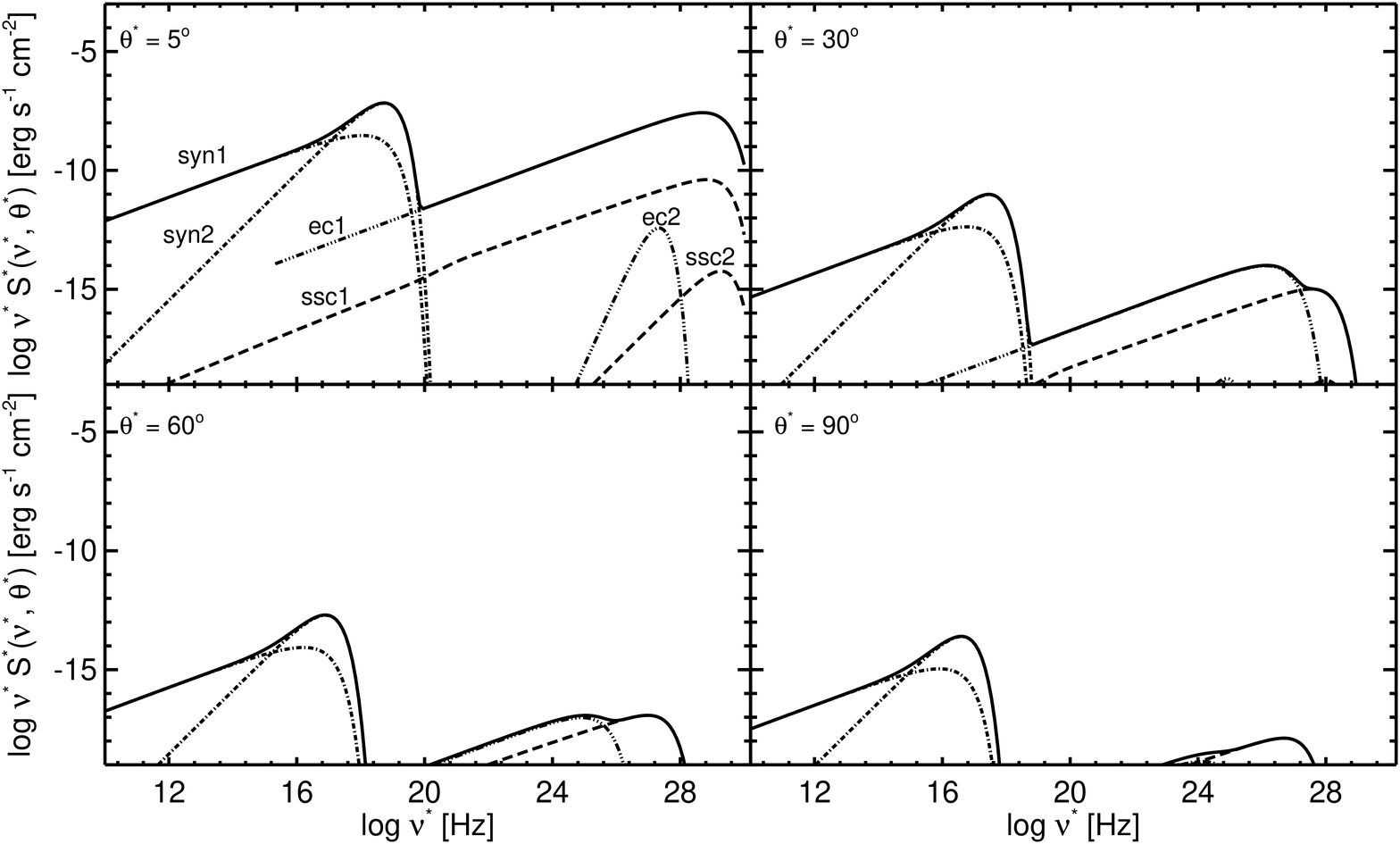}
  \caption{Predicted spectral energy distributions of the radiation
    generated by an AR boundary layer ($\Gamma_{\rm c}=10$,
    $\Gamma_{{\rm j},2}=20$) of a kpc-scale jet assuming $\gamma_{\rm
      eq}=10^8$, for observing angles
    $\theta^\ast=5^\circ, 30^\circ, 60^\circ$ and $90^\circ$. Indices
    1 and 2 denote the spectral components associated to the power-law
    isotropic energy distribution and to the monoenergetic part,
    respectively. Absorption of VHE gamma-rays during the propagation
    to the observer is neglected.
    \label{fig:SED_AR-thetas}}
\end{figure}

\begin{figure}
\plotone{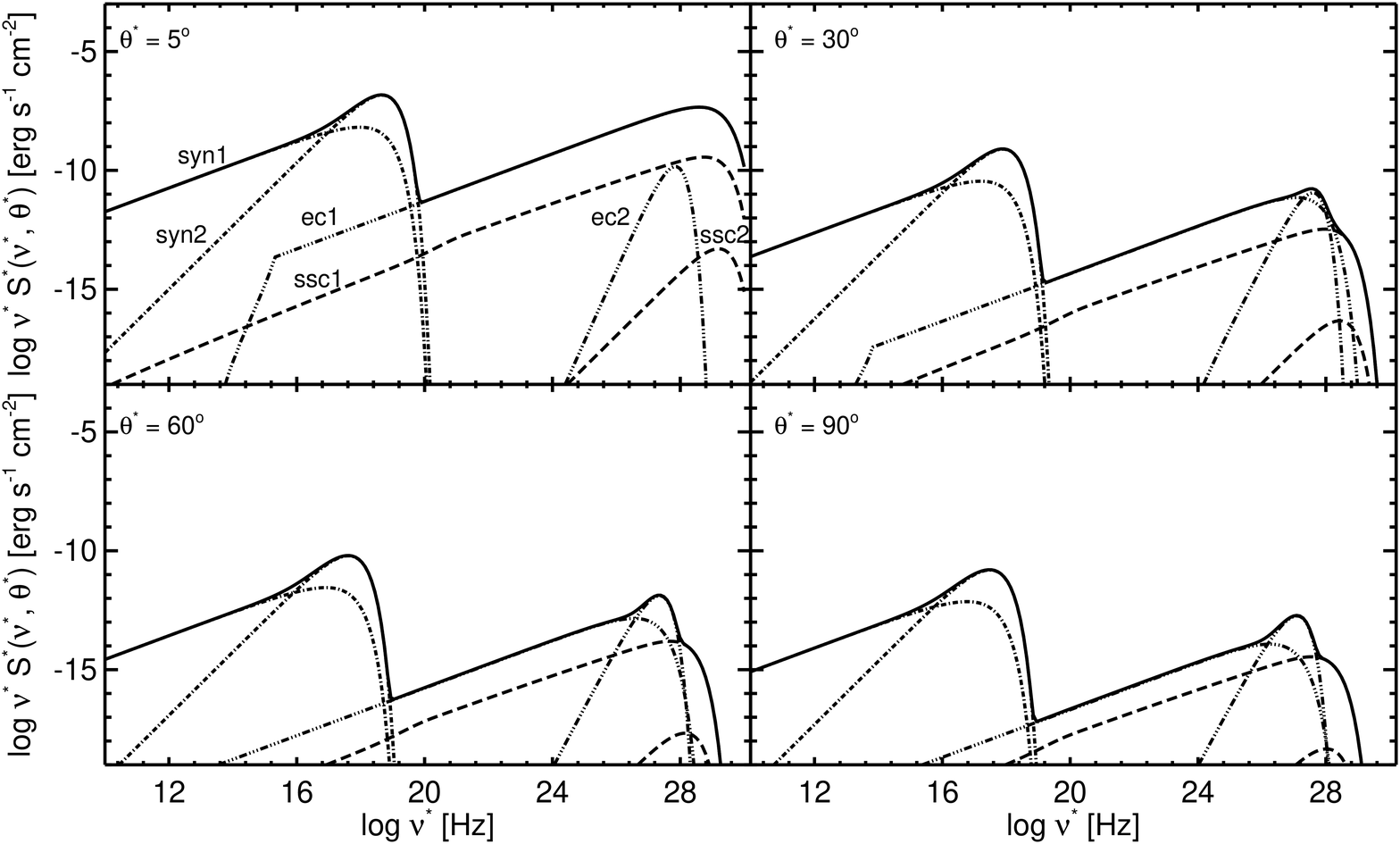}
  \caption{Same as Fig.\ref{fig:SED_AR-thetas} but for a standard
    boundary layer ($\Gamma_{\rm c}=10$, $\Gamma_{{\rm j},1}=1$).
    \label{fig:SED_uniform-thetas}}
\end{figure}

Considering specifically the flux density produced by anomalous layers
(Fig.~\ref{fig:SED_AR-thetas}), we note two main differences with the
standard case studied by SO02. First, the separation of the SSC and EC
peaks corresponding to IC emission of the power-law distribution of
the electrons (Eq.~\ref{eq:n(gamma)}) is larger in the former case.
For jets with AR layers the frequency separation between the EC1 and
SSC1 peaks grows with increasing viewing angle from less than one
order of magnitude ($\theta^\ast\sim 5^\circ$) to more than two orders
($\theta^\ast\sim 90^\circ$). In case of jets endowed by standard
layers (Fig.~\ref{fig:SED_uniform-thetas}), the frequency separation
between the EC1 and SSC1 peaks also grows with the viewing angle, but
much less than in the previous case: the frequency separation between
the two peaks is smaller than a factor 8 at $\theta^\ast=90^\circ$. If
a sufficiently large number of TeV photons were detected, this
difference could be used to distinguish between jets with standard and
anomalous layers. However, we repeat here the cautionary note of SO02:
the spectral energy distributions (SEDs) computed in this paper
correspond to the most optimistic scenario, with a highly relativistic
jet spine and very efficient acceleration creating electrons with
large Lorentz factors up to $\gamma_{\rm eq}\sim 10^8$. Either smaller
values of $\Gamma_{\rm c}$ or $\gamma_{\rm eq}$ would reduce the
observed very high energy gamma-ray flux and shift the corresponding
peaks to lower energies.

The second most important difference between AR and standard radiating
boundary layers is that the exponential decay of the synchrotron spectrum
is shifted towards smaller frequencies (delving deeper into the very
soft X-ray regime). This has a big impact in the effective X-ray
spectral index as we will discuss below.

\begin{figure}
\plotone{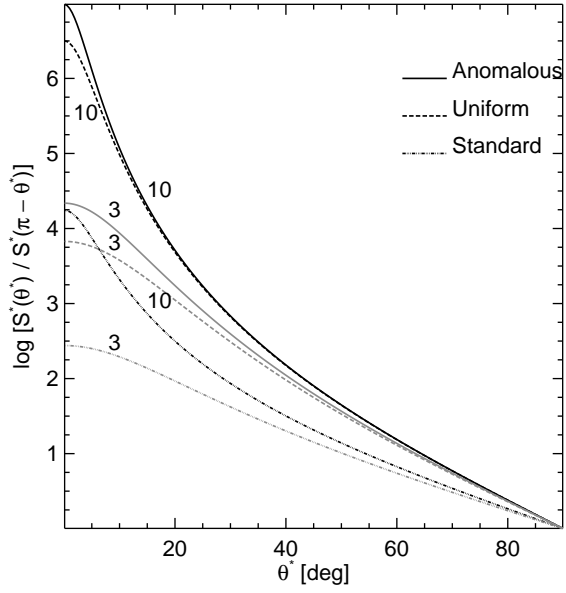}
\caption{Jet-to-counterjet flux density ratio,
  $S^{\ast}(\theta^{\ast}) / S^{\ast}(\pi - \theta^{\ast})$ at radio
  frequencies, as a function of the viewing angle
  $\theta^{\ast}$. Black and grey lines correspond to values
  $\Gamma_{\rm c}=10$ and $\Gamma_{\rm c}=3$, respectively --the
  values of $\Gamma_{\rm c}$ are provided near the respective
  curves--. The black (grey) solid line corresponds to the radiation
  from an anomalous boundary shear layer with the radial profile of
  Eq.~\ref{eq:Gamma} and $\Gamma_{{\rm j},1}=15$ ($\Gamma_{{\rm
      j},1}=4.5$). The dashed-doted line corresponds to the radiation
  from a monotonic boundary layer with the radial profile of
  Eq.~\ref{eq:Gamma} and $\Gamma_{{\rm j},2}=1$. The dashed line
  corresponds to the model with a uniform boundary layer with a
  Lorentz factor $\Gamma_{{\rm j},3}=10$. \label{fig:jet2counterjet}}
\end{figure}

%
%
As noted by \cite{kom90} and confirmed by SO02, extragalactic jets
with a monotonic boundary layer may show a jet-to-counterjet
radio-to-optical brightness ratio ($S^{\ast}_{\rm syn} (\theta^{\ast})
/ S^{\ast}_{\rm syn} (\pi - \theta^{\ast})$) which is smaller than if
the jet were uniform. We find that such an assertion has to be
modified for jets limited by AR-boundary layers. In the latter case,
at small viewing angles (namely, $\theta^\ast\simlt 5^\circ$ for the
considered parametrization) the jet/counterjet asymmetry is $\sim 2 -
3$ times larger than the one corresponding to a uniform jet with the
same kinetic power (Fig.~\ref{fig:jet2counterjet}). At larger viewing
angles, the brightness asymmetry turns out to be almost
indistinguishable from the case of an equivalent uniform jet (see in
Fig.~\ref{fig:jet2counterjet} how the solid and dashed lines
practically overlap each other for $\theta^\ast \simgt 20^\circ$ if
$\Gamma_{\rm c}=10$ or for $\theta^\ast \simgt 50^\circ$ in case
$\Gamma_{\rm c}=3$). Certainly, since the brightness asymmetry is a
result of the Doppler boosting (hiding) of the jet (counterjet), the
smaller is the value of the maximum Lorentz factor in the anomalous
shear layer ($\Gamma_{{\rm j},2}$), the smaller will be the maximum
asymmetry at small $\theta^\ast$. If the Lorentz factor of the jet
spine is small, the jet-to-counterjet brightness ratio reduces in
absolute value (see the grey lines corresponding to $\Gamma_{\rm
  c}=3$, in Fig.~\ref{fig:jet2counterjet}).%
\footnote{The relative increase of the Lorentz factor in an anomalous
  shear layer is rather independent on the Lorentz factor of the jet
  (left state). Therefore, since we have chosen a parametrization for
  our prototype jet in which $\Gamma_{{\rm j},2}/\Gamma_{\rm c}=1.5$,
  for the case of $\Gamma_{\rm c}=3$ we take $\Gamma_{{\rm
      j},2}=4.5$.}
However, even for the relatively small value of $\Gamma_{\rm c}=3$ the
jet-to-counterjet brightness ratio of the anomalous shear layer case
is larger than the one corresponding to the standard case at
$\Gamma_{\rm c}=10$ independent of the viewing angle (e.g., at
$\theta^\ast=0^\circ$, $\log{\left[S^{\ast}_{\rm syn} (0) /
    S^{\ast}_{\rm syn} (\pi)\right]} \simeq 4.35$ for the AR model,
while $\log{\left[S^{\ast}_{\rm syn} (0) / S^{\ast}_{\rm syn}
    (\pi)\right]} \simeq 4.25$). If there were a handle on the value
of the Lorentz factor of the jet, this difference might allow us to
discriminate observationally between jets with anomalous and standard
shear layers. However, given the similarity of the jet-to-counterjet
brightness ratio between jets with anomalous layers and jets with {\it
  top-hat} profiles (uniform), it might be practically impossible to
disentangle from this unique value whether the jet is shielded by an
anomalous layer or by no layer at all.

Considering the results obtained for standard boundary layers, SO02
concluded that the mildly relativistic velocities inferred from the
observed brightness asymmetries at tens of kiloparsec scales, may
correspond to a slower boundary layer and not necessarily to the faster
jet core, which might be highly relativistic at the observed
distances. If anomalous shear layers may happen in actual jets,
slower spines are preferred and relatively moderate maximum shear
layer Lorentz factors might also be invoked to explain the observed
jet-to-counterjet asymmetries.

%
%
In our models, the X-ray radiated power of kpc-scale jets bounded by
standard layers is dominated by the high-energy electron bump spectral
component, implying that the resulting synchrotron X-ray flux lies
above the extrapolated radio-to-optical continuum (see SO02 and
Figs.~\ref{fig:SED_AR-SO}, \ref{fig:SED_uniform-thetas}). This is also
true for jets endowed by anomalous boundary layers (see
Fig.~\ref{fig:SED_AR-thetas} at $\nu^\ast \simeq 10^{17}\,$Hz). For
both types of layers, the spectral slope at X-ray frequencies may be
rather different from that of the power-law at radio-to-optical
frequencies, and the difference depends also on the viewing
angle. This implies that the effective X-ray spectral index
$\alpha_{\rm X, eff}(\theta^\ast)$ computed between $h\nu^\ast_1 =
1\,$keV and $h\nu^\ast_2 = 5\,$keV as
\begin{equation}
  \alpha_{\rm X, eff}(\theta^{\ast}) = { \log \left[ S^{\ast}_{\rm X}
      (\nu^{\ast}_1, \theta^{\ast}) \, / \, S^{\ast}_{\rm X} (\nu^{\ast}_2,
      \theta^{\ast}) \right] \over \log \left[ \nu^{\ast}_2 \, / \,
      \nu^{\ast}_1 \right] } ,
\label{eq:alphaX}
\end{equation}
can be significantly distinct. We point out that $S^\ast_{\rm
  X}(\nu^{\ast}, \theta^{\ast})$ is computed in this paper as the
value of the flux density calculated using Eq.~\ref{eq:S(nu,theta)}
instead of SO02 who only take $S^\ast_{\rm X}(\nu^{\ast},
\theta^{\ast}) \propto \int r dr {\cal D}^2(r,\theta^\ast)
R(x^\ast_{\rm eq})$, with $R(x^\ast_{\rm eq})$ given by
Eq.~\ref{eq:R(x)}, and $x^\ast_{\rm eq}= \nu^\ast/c_1\gamma^2_{\rm eq}
{\cal D}(r,\theta^\ast)$.

For highly relativistic jet cores ($\Gamma_{\rm c}=10$),
Fig.~\ref{fig:alphaX} (black lines) shows that the most noticeable
difference between jets with anomalous or uniform shear layers and
jets with standard limiting boundaries, is the fact that the effective
X-ray spectral index of the jet $\alpha_{\rm X, eff}^{\rm j}$ is
larger than the corresponding to the counterjet $\alpha_{\rm X,
  eff}^{\rm cj}$ for jet inclinations $\theta^\ast \simgt
65^\circ$. For standard boundary layer jets $\alpha_{\rm X, eff}^{\rm
  j} \leq \alpha_{\rm X, eff}^{\rm cj}$, $\forall \theta^\ast$ holds
(i.e., the counter jet has a steeper X-ray continuum as compared to
the jet spectrum). Furthermore, jets with uniform and AR layers show a
much larger%
\footnote{Particularly, in the range $\theta^\ast \in
  [60^\circ,90^\circ]$, $\alpha_{\rm X, eff, (Uniform, AR)}^{\rm j}$
  is 6 to 7 times larger than $\alpha_{\rm X, eff, standard}^{\rm
    j}$.}
jet and counter jet effective X-ray spectral index than jets with
standard layers if the inclination is $\theta^\ast \simgt 30^\circ$.
Indeed, $\alpha_{\rm X, eff}^{\rm j}$ in the former two types of
boundary layers is even larger than the $\alpha_{\rm X, eff}^{\rm cj}$
corresponding to jets with monotonic layers for viewing angles
$\theta^\ast \simgt 30^\circ$.

The large values of the effective X-ray spectral index in jets with
uniform and AR layers is due to the fact that the decay of the
spectrum after the synchrotron peak crosses the X-ray band and moves
towards optical frequencies for inclinations $\theta^\ast \simgt
60^\circ$ (Fig.~\ref{fig:SED_AR-thetas}). Consistently, the flux
density at 5\,keV decays abruptly and, thus $\alpha_{\rm X, eff}$
grows.

In case of jets with more moderate spine velocities ($\Gamma_{\rm
  c}=3$; Fig.~\ref{fig:alphaX}, grey lines), there is no viewing angle
for which $\alpha_{\rm X, eff}^{\rm j} > \alpha_{\rm X, eff}^{\rm
  cj}$, independent on the shear layer model. However, it remains true
that the counterjet X-ray spectral index of jets bearing uniform or AR
layers is much larger than the one corresponding to standard boundary
layer jets (there is a factor of 2 to 4 difference between them). 

Taking together the results for models with jet spines flowing at
$\Gamma_{\rm c}=3$ and 10, it turns out that the range of variability
of both $\alpha_{\rm X, eff}^{\rm j}$ and $\alpha_{\rm X, eff}^{\rm
  cj}$ is larger for jets flanked by uniform and AR layers than for
jets with standard boundary layers. Interestingly, the large-scale
X-ray emission observed from several radio-loud AGNs exhibits rather
different spectral characteristics. For instance, X-ray spectra of the
known jets in quasars are very flat (e.g., $\alpha_{\rm X} \sim 0.23$
for 3C 207, \citealp{bru01}; $\alpha_{\rm X} \sim 0.5$ for PKS 1127,
\citealp{sie02}; $\alpha_{\rm X} \sim 0.8$ for 3C 273 and PKS
0637). The low effective spectral index is consistent with small jet
inclinations (c.f. Fig.~\ref{fig:alphaX}, inset) independent of the
shear layer model adopted. On the other hand, jets in radio galaxies
tend to display relatively steep X-ray spectra, $\alpha_{\rm X}>1.0 -
1.5$ (\citealt{hac01} for 3C 66B, \citealt{wor01} for B2 0206 and B2
0755).  These values correspond to large inclination angles on
Fig.~\ref{fig:alphaX}, and/or lower jet Lorentz factors for jets with
standard boundary layers. But, if uniform or AR emitting layers are
considered, the viewing angles needed to account for such values of
the X-ray spectral index are smaller ($\theta^\ast \in [20^\circ,
25^\circ]$ if $\Gamma_{\rm c}=10$, or $\theta^\ast \in [40^\circ,
55^\circ]$ if $\Gamma_{\rm c}=3$). Very recently, \cite{hac07} have
reported values of the X-ray spectral index $\alpha_{\rm X}>2.5$ at
distances of $\simeq 4\,$kpc from the nucleus in Cen A. These large
values of $\alpha_{\rm X}$ are not within reach of jet models flanked
by standard layers (the maximum of $\alpha_{\rm X, eff}^{\rm j}$ is
1.84 at $\theta^\ast=90^\circ$ for the model with $\Gamma_{\rm
  c}=10$). Nonetheless, jets with uniform or AR boundaries may yield
$\alpha_{\rm X, eff}^{\rm j}>2.5$ for $\theta^\ast>32^\circ$
($\theta^\ast>59^\circ$) if $\Gamma_{\rm c}=10$ ($\Gamma_{\rm
  c}=3$). Indeed, considering that the inclination of Cen A is $\sim
60^\circ - 70^\circ$ \citep{SDK94}, our results suggest that, if a
uniform or an AR emitting boundary layer were responsible for the
emission of Cen A at distances from the nucleus $\simgt 4\,$kpc, small
values of the spine Lorentz factor ($\Gamma_{\rm c}\simeq 3$) are
preferred to explain the large values of $\alpha_{\rm X}$ observed at
such scales.

Although very recently deep {\it Chandra} of observations of Cen A
\citep{hac07} have provided us with spectra of some counterjet X-ray
features, this is not the usual case for large scale
counterjets. There are, however, some estimates of the lower limit of
the X-ray brightness asymmetry, like e.g., for the 3C 66B ($> 25$,
\citealt{hac01}), Pictor A ($> 15$, \citealt{wil01}; $6^{+12}_{-2}$
\citealt{HC05}), or PKS 1127-145 ($> 5$, \citealt{sie02}).  Our model
of AR or uniform radiating boundary layer tends to favor larger X-ray
brightness asymmetries between the jet and the counterjet. The reason
being that the X-ray emission of the jet is dominated by the
synchrotron radiation from the monoenergetic, hard ($\gamma=10^8$)
electron component, while the counterjet emission is dominated by the
EC radiation of the power-law part of the electron spectrum.

\section{Discussion}
\label{sec:discussion}

SO02 chose the parameters of their radiating boundary layers such that
they exhibit clearly the main characteristics of their model. We do
not repeat here their discussion. Instead, we focus on the discussion
of the choice of the new parameters of our model.

SO02 anticipated that the form of the $\Gamma(r)$ radial profile and
of the related spatial variation of the acceleration efficiency can
significantly influence the beaming pattern and the intensity of the
boundary layer emission. We actually have confirmed this point by
replacing the monotonically decaying profile of $\Gamma(r)$ by an
idealization of the Lorentz factor profile that results from AR
boundaries. This results in a different relative normalization of the
SSC and EC components presented in
Figs.~\ref{fig:SED_AR-SO}-\ref{fig:SED_uniform-thetas}.

\begin{figure}
\epsscale{0.85}
\plotone{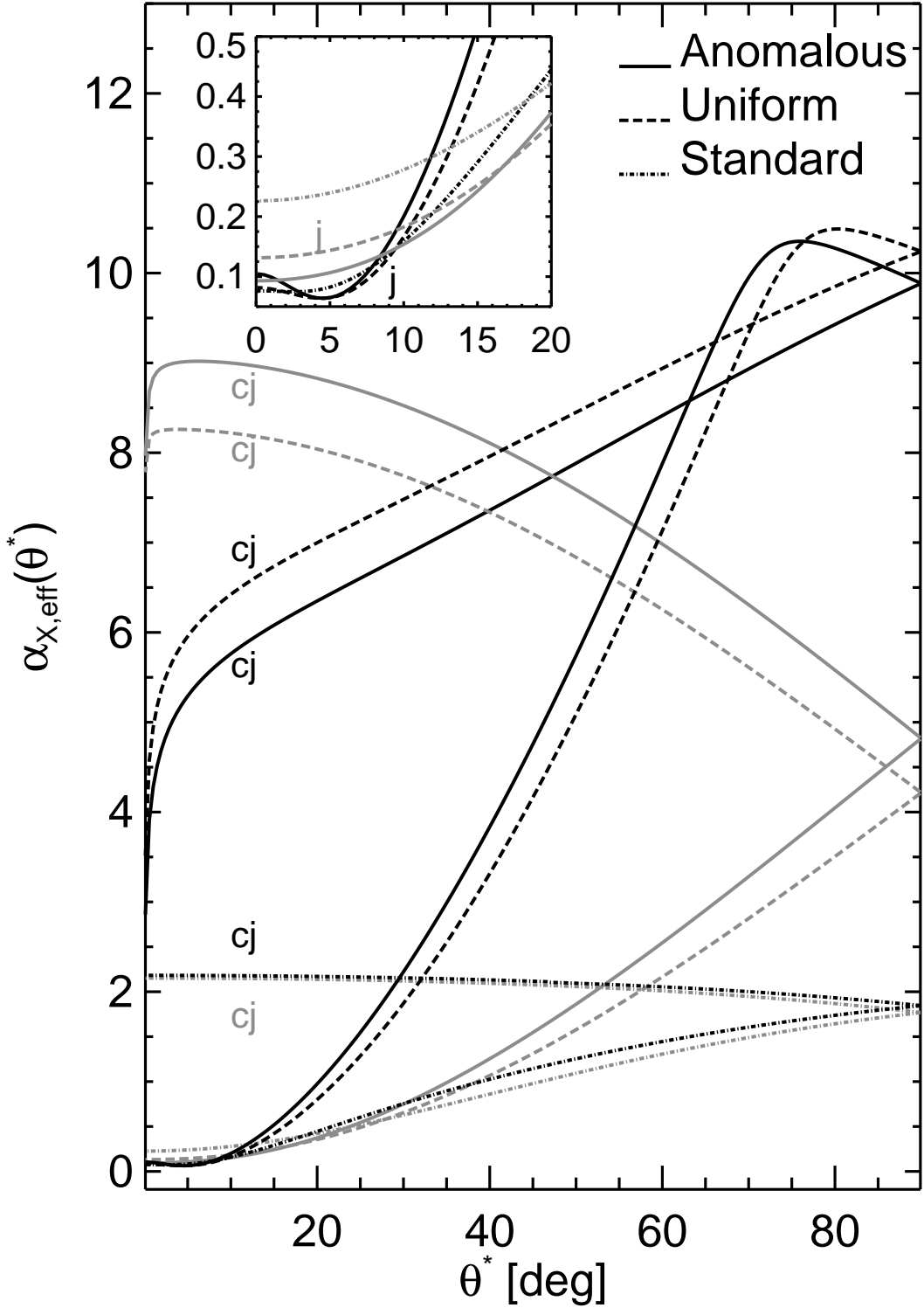}
\caption{Expected effective X-ray spectral index of a jet (j) and of a
  counterjet (cj) boundary layer emission, $\alpha_{\rm X,
    eff}(\theta^\ast)$, as a function of the viewing angle
  $\theta^\ast$. Solid, dash-dotted and dashed lines correspond to
  jets with anomalous, standard and uniform shear layers, respectively
  (Eq.~\ref{eq:Gamma}). Black (grey) lines correspond to jets whose
  core moves with a Lorentz factor $\Gamma_{\rm c}=10$ ($\Gamma_{\rm
    c}=3$). The inset shows a zoom of the jet $\alpha_{\rm X, eff}$
  for the smallest viewing angles $\theta^\ast$. \label{fig:alphaX}}
\end{figure}

The uniform layer model could be considered as a limiting case of AR
layers when $\Gamma_{\rm j,i} \rightarrow \Gamma_{\rm c}$. Thus, when
looking at the plots of the jet-to-counterjet ratio
(Fig.~\ref{fig:jet2counterjet}) or the X-ray effective spectral index
(Fig.~\ref{fig:alphaX}), any model including an AR layer with
$\Gamma_{\rm j,i} \in [10,15]$ (if $\Gamma_{\rm c}=10$; black lines in
Figs.~\ref{fig:jet2counterjet} and \ref{fig:alphaX}) or $\Gamma_{\rm
  j,i} \in [3,4.5]$ (if $\Gamma_{\rm c}=3$; grey lines in
Figs.~\ref{fig:jet2counterjet} and \ref{fig:alphaX}), would display a
graph located between the boundaries set by the solid and dashed lines
of the corresponding figures. Since the area enclosed by these two
line types (solid and dashed) is rather small, it turns out that our
choice of the value of $\Gamma_{\rm j,i}=1.5\Gamma_{\rm c}$ does not
significantly influence the results. More precisely, if for the
considered values of $\Gamma_{\rm c}$ (3 or 10) we would have picked
up any other value of $\Gamma_{\rm j,i} \in [\Gamma_{\rm c},
1.5\Gamma_{\rm c}]$, neither the value of the jet-to-counterjet
brightness ratio, nor $\alpha_{\rm X}$, nor even the spectrum would
have changed appreciably. Since, keeping all other physical conditions
fixed, a decrease of ${\tilde \beta}_{_{\rm L}}$ yields smaller values of
$\Gamma_{\rm j,i}$, the results we obtain shall be qualitatively valid
both for magnetized or non-magnetized jets.

The jet-to-counterjet effective X-ray spectral index asymmetry and its
angular dependence (Fig.~\ref{fig:alphaX}) depends on which of the two
spectral components at X-ray frequencies dominates: synchrotron
radiation of the monoenergetic, high-energy electron component or the
EC radiation from low energy electrons ($\gamma \sim 100$). In its
turn, the dominance of any of these two processes depends on the exact
shape of the electron spectrum at the highest energies, on the viewing
angle, on the jet Lorentz factor and on
$\Gamma(r)$. Figure~\ref{fig:alphaX} serves as an example of the
comparative behavior of different boundary layer models (in which all
the parameters are the same except $\Gamma(r)$).

\subsection{Summary}
\label{sec:summary}

In this work, we compare the observational signatures imprinted by
different kinematic models of shear layers of kiloparsec scale jets on
the radiation of ultrarelativistic electrons accelerated at such
boundary layers. The (simple) radiative model of a jet boundary layer
follows that of SO02. Alternative (more sophisticated) models to
compute the boundary layer radiation from multizone models (see, e.g.,
\citealt{MA04,MA05,MA07}) will be considered elsewhere. The most
important difference among distinct kinematic models is the radial
profile of the Lorentz factor $\Gamma(r)$ across the shear layer.  We
have considered three different profiles. The first one is the usually
assumed for large scale jets, namely, a monotonically decaying Lorentz
factor profile (as in SO02). The second one is motivated by the
results of \cite{AR06} and \cite{Mizuetal08}, namely, a non-monotonic
profile where the Lorentz factor reaches a maximum at the limit
between the jet and the external medium (AR layer). For completeness,
the case of a uniform jet with a sharp edge is considered. The last
case can be regarded as a limit of the AR profile when the maximum
Lorentz factor in the boundary layer equals the Lorentz factor of the
jet spine. Along the way, we have conveniently developed the results
of \cite{AR06} and \cite{Mizuetal08} to include the effects of
dynamically important magnetic fields in the parameter range which is
adequate for large scale jets. From such development we have obtained
and idealized kinematic model of anomalous boundary layers in
magnetized extragalactic jets.

Not surprisingly, we find that the jet-to-counter jet brightness ratio
at radio frequencies is larger for jets flanked by AR layers than for
uniform or standard layers with the same jet core Lorentz factor. This
results from the fact that the differences in the radial profile of
the velocity field determine the beaming pattern of the boundary layer
emission. Thereby, this effect has to be taken into account when the
jet bulk Lorentz factor is inferred from the jet-to-counterjet
brightness asymmetry.

The differences between AR and uniform boundary layer jets are small
in the framework of our simple model. In practice, these small
differences might render their observational distinction very
difficult. Comparing these two models with standard radiating boundary
layers the differences are much larger. Several independent clues can
be used to distinguish (observationally) among them:
\begin{enumerate}
\item If a sufficiently large number of TeV photons were detected, it
  would be possible to distinguish between jets with standard and
  anomalous layers by looking at the separation of the SSC and EC
  peaks. The separation between the two peaks is larger for jets with
  AR or uniform layers.
\item A large jet-to-counter jet brightness ratio (about two orders of
  magnitude larger than in standard boundary layer jets) is expected
  for jets bounded by AR or uniform layers.
\item For large viewing angles the effective X-ray spectral index is much
  larger for AR or uniform jets than for jets with standard boundary
  layers. 
\item There can exist an inversion of the jet and counterjet X-ray
  spectral indices for jets seen at large viewing angles. However, for
  most inclination angles and moderate bulk Lorentz factors of the jet
  core, the effective X-ray spectral index of the jet is much smaller
  than that of the counterjet. Indeed, if there is a hint on the bulk
  Lorentz factor of the jet (e.g., because superluminal proper motions
  are detected), the jet-to-counterjet spectral X-ray index ratio may
  tell us whether the radiating layer is uniform, standard or AR. We
  note that for large Lorentz factors of the jet core ($\Gamma_{\rm c}
  \simlt 8$), the largest ratio $\alpha_{\rm X, eff}^{\rm j} /
  \alpha_{\rm X, eff}^{\rm cj}$ corresponds to jets with uniform
  layers, whilst the smaller one should be identified with standard
  boundary layers. For more moderate core Lorentz factor, the ratio
  $\alpha_{\rm X, eff}^{\rm j} / \alpha_{\rm X, eff}^{\rm cj}$ is
  larger for jets flanked by AR layers than for jets with uniform or
  standard layers. In spite of these facts, the differences in the
  effective spectral X-ray index between models with uniform and with
  AR layers are rather small and, they might render their
  observational distinction very difficult unless a very careful
  analysis of the observational data is performed.
\end{enumerate}

\acknowledgements
  We kindly acknowledge the work of Jerome Ferrand during his
  internship at the Departamento de Astronom\'{\i}a y
  Astrof\'{\i}sica.  We also thank L. Stawarz for very useful
  discussions and helpful hints. MAA is a Ram\'on y Cajal Fellow of
  the Spanish Ministry of Education and Science. He also acknowledges
  partial support from the Spanish Ministry of Education and Science
  (AYA2004-08067-C03-C01, AYA2007-67626-C03-01, CSD2007-00050). PM is
  at the University of Valencia with a European Union Marie Curie
  Incoming International Fellowship (MEIF-CT-2005-021603). The authors
  thankfully acknowledge the computer resources, technical expertise
  and assistance provided by the Barcelona Supercomputing Center -
  Centro Nacional de Supercomputaci\'on.


\clearpage

\clearpage

\clearpage

\clearpage

\clearpage

\clearpage

\clearpage

\clearpage

\clearpage

\end{document}